\documentclass[aps,prl,showkeys,showpacs,preprint,groupedaddress]{revtex4}
\usepackage{graphicx}

\begin{document}

\title{Quantum Theory finally reconciled with Special Relativity}

\author{Daniele Tommasini}

\affiliation{Departamento de F\'\i sica Aplicada, \'Area de
F\'\i sica Te\'orica, Universidad de Vigo, 32004 Ourense, Spain}

\email[]{daniele@uvigo.es}

\date{\today}

\begin{abstract}
In 1935 Einstein, Podolsky and Rosen (EPR) pointed out that
Quantum Mechanics apparently implied some mysterious,
instantaneous action at a distance. This paradox is supposed to be
related to the probabilistic nature of the theory, but since
deterministic alternatives involving ``Hidden Variables" hardly
agree with the experiments, the scientific community is now
accepting this ``quantum nonlocality" as if it were a reality.
However, I have argued recently that Quantum Electrodynamics is
free from the EPR paradox, due to an indetermination on the number
of the unobserved ``soft photons" that can be present in any step
of any experiment. Here, I will provide a more general proof,
based on an approach to the ``problem of measurement" that implies
the full reconciliation of Quantum Field Theory with Special
Relativity. I will then conclude with some considerations on the
interpretation of the Quantum Theory itself.
\end{abstract}



\maketitle

In 1935 Einstein, Podolsky and Rosen (EPR) \cite{EPR} pointed out
that Quantum Mechanics apparently implied some mysterious,
instantaneous action at a distance. This was considered a paradox,
since it made the Quantum Theory incompatible with Special
Relativity, suggesting the need for a more fundamental description
of Nature. For a long time, many physicists have tried to build a
causal local, deterministic theory, in the belief that the
probabilistic behavior of the microscopic world was only due to
the non observation of some Hidden Variables. Apart from the
difficulty in elaborating a satisfactory model, such a project is
now considered to be undermined by the results of several
experiments that have been performed in the last 20 years
\cite{Aspect,Zeilinger}. This has lead a large part of the
scientific community to accept the Quantum Theory, including its
supposed nonlocality, that is even used as the base for
Teleportation and the current theories for Quantum Information and
Quantum Computers. However, in the last few decades a very
beautiful and successful description of the fundamental
interactions has been achieved in terms of Quantum Field Theories
(QFTs) \cite{WeinbookI,WeinbookII}. Recently, I have argued that
the most popular of these theories, Quantum Electrodynamics (QED),
is free from the EPR paradox, due to an indetermination on the
number of the unobserved ``soft photons" (i.e. photons that carry
a very low energy) that can be present in any step of any
experiment \cite{QEDEPR}. Here, I will provide a more general
proof, based on an approach to the ``problem of measurement" that
implies the full reconciliation of QFT with Special Relativity. I
will then conclude with some considerations on the interpretation
of the Quantum Theory itself.

Let me first briefly describe a typical, ideal EPR experiment,
following Chap.˜ 17 of Ref.˜ \cite{BJ}. Two particles A and B, say
two electrons \footnote{The following discussion can be
generalized, e.g. to include the case, that has been studied in
actual experiments \cite{Aspect,Zeilinger}, where the particles
under measurement are two (or more) photons \cite{QEDEPR}.}, are
emitted in coincidence by a source and travel in opposite
directions. Far from the production point, some conserved
observable, say the $z$-component of the angular momentum (spin),
is measured on both of them.

To fix the ideas, suppose that we know that the total angular
momentum is zero, and that each of the two particles is created
without a definite value for the spin component $S_z$. According
to ordinary Quantum Mechanics, the measurement of the spin
$S_z(A)$ of the electron A that is produced in a given single
event will yield either the value $+\hbar/2$, or the value
$-\hbar/2$. If the experiment is repeated a great number of times,
each of these two results will appear with a frequency that will
be an approximation for the corresponding probability. For
instance, if the initial state were completely ``unpolarized",
these two probabilities would be equal to $1/2$. Now, let me
consider again a single event, and measure the spin $S_z(A)$ of
particle A far from the source, obtaining -say- the value
$+\hbar/2$. According to the usual interpretation \cite{BJ}, the
conservation of the total angular momentum would then force
particle B to get the definite value $-\hbar/2$ of $S_z(B)$. In
other words, particle B would immediately change its state,
``collapsing" in the eigenstate $\vert-\hbar/2>$ of $S_z(B)$, as a
mere consequence of the measurement performed on the distant
particle A! Such a situation is called a violation of ``local
realism", and shows an evident contradiction with Relativity, that
forbids any action to propagate faster than light. This is the
Einstein-Podolsky-Rosen paradox.

In practice, what can be done to test locality according to the
previous ideas? It is clear that it is not sufficient to measure
the spins of the two electrons in a single event and check that
the spin of B takes a value which is opposite to that of A. In a
single event, this could just be due to a casuality. For the test
to be significant, in our simple example {\it it seems that} we
have to check two things: 1) each of the two electrons should not
be prepared in a $S_z$ eigenstate before the measurement (this
condition can be guaranteed a priori by the experimental settings,
and can be controlled a posteriori by verifying that different
single measurements do not always give the same result for the
individual spins); 2) the two spins should {\it always} give
opposite values, whatever event is considered. In other words, we
have to check that the two spins are ``maximally correlated"
\footnote{For simplicity, I will consider only such a case, that
corresponds to our two electron system. Any complication of such a
scheme is not relevant for the present discussion.}. In fact, this
seems to be a consequence of Quantum Mechanics. For a long time,
the experimental study of the correlations amongst the components
of the spins of A and B along two arbitrary (in general different)
directions in space, was also (erroneously) believed to allow to
prove (or to discard) ``quantum nonlocality" \footnote{Actually, a
more elaborated test was invented by Bell \cite{Bell} that allowed
to evaluate just some combinations of the correlations. A set of
inequalities made out of them was then argued to be satisfied by
Hidden Variable Theories and violated by Quantum Mechanics.}.

Let me now come back to the paradox, and make some historical
remarks. Since the authors of Ref.˜ \cite{EPR} considered that
local realism was a necessary ingredient of any reasonable
physical description of Nature, they judged that a new theory was
needed. Einstein also thought that the problem was due to the
intrinsically probabilistic description of Quantum Mechanics, and
hoped that a solution would eventually be found in Hidden Variable
Theories, that would also allow for a deterministic description of
Nature (``God does not play dice", he used to say). Was he right?

In the last few decades, there has been increasing agreement
within the scientific community that he was wrong. This conclusion
seemed to be justified by the actual realization of a series of
experiments measuring the statistical correlations between the
spin/polarizations of the two (or more) particles A, B, that were
produced in coincidence in an EPR experiment. The data were
considered to be incompatible with Hidden Variables Theories,
while they agreed with the prediction of Quantum Mechanics. Since
Einstein himself used to relate locality with determinism, it is
perhaps not surprising that these results were also interpreted
(erroneously!) as an experimental evidence of ``quantum
nonlocality". This conclusion was then transformed in the
so-called ``Bell theorem". As a consequence, instead of being
concerned with that implicit violation of Special Relativity that
was apparently implied by Quantum Mechanics, several physicists
have decided to accept the EPR paradox, i.e. that mysterious
``quantum nonlocality", as if it were a real characteristic of
Nature itself. It is then generally believed that some mysterious
action at a distance continues to link the fate of distant
particles that have been produced in coincidence in the past. Such
a supposed ``phenomenon" is also called {\it entanglement}, and it
is also believed to allow for the {\it teleportation} of the state
of a particle A to another distant particle B. Enormous efforts in
using the EPR paradox to build Quantum Information Theories and
models for Quantum Computers have then been performed. To justify
the strong violation of intuition and, what is a more serious
problem, of Relativity, that is implied by such applications, it
is common to cite some old sentences of authorities such as
Richard P. Feynman (``I think I can safely say that nobody today
understands Quantum Physics") or Roger Penrose (Quantum Theory
``makes absolutely no sense"). Ironically, Feynman himself is also
one of the main contributors to the construction of the QFTs that
allow for the solution of the paradox and the recovery of local
realism and of a good deal of intuition.

In fact, the above argument leading to the EPR paradox is based on
ordinary Quantum Mechanics, with two hidden assumptions: i) before
the measurement is done, only the particles A and B are supposed
to exist, i.e. no additional particle is allowed to appear in
coincidence with them (this implies that, when the spin of A is
measured and fixed to a definite value, it can only be compensated
by an opposite value of the spin of B in order to conserve the
zero total angular momentum); ii) the measurement process is
supposed to modify the state of particle A, forcing it to
`collapse" ìnto an eigenstate of the observed magnitude (e.g.
angular momentum), {\it and the actual modifications induced in
the measuring apparatus are not considered to be relevant.}

However, {\it the old Quantum Mechanics approach is not correct in
QFT, and these two assumptions turn out to be wrong.}

In Ref.˜ \cite{QEDEPR}, I have already disproved assumption i).
That argument was related to a basic characteristic of QFTs: {\it
they predict a non vanishing probability for any process that does
not violate any fundamental symmetry}. It is important to point
out that this corresponds perfectly to the behavior observed in
the High Energy Physics experiments: for instance, it is well
known that an electron-positron collision can produce any result
(each with its own probability) that is allowed by the available
energy and the conservation of momentum, angular momentum,
electric and color charge, etc. It is so rare to find an
``accidental cancellation" for the rate of an allowed process,
that such a case would be considered as a hint for some new
symmetry forbidding that channel \footnote{This is the case e.g.˜
of Baryon and Lepton Numbers conservations, that are often thought
to be due to some Grand Unification.}. In other words, all the new
particles that can be created without violating the universal
conservation laws can actually be produced, and any definite
process involving the creation of a particular set of particles
has his corresponding amplitude of probability, that can
eventually be computed approximately by drawing the relevant
Feynman diagrams (when perturbation theory is applicable). In
particular, since photons have zero rest mass, their energy can be
arbitrarily low. Since they also have zero charge and color, we
can conclude that {\it an arbitrary number of ``soft" photons,
with low enough total energy, can always be created in coincidence
with any physical process!} It is important to point out that
these soft photons can exist even if they are not observed: not
only are they not looked for, but they would also easily escape
detection anyway, due to their very low energy.

Incidentally, this result by itself will imply very important
consequences for the Theory of Measurement. For instance,
according to the usual postulates, the measurement of an
observable in a system that is previously in one of its
eigenstates will leave the system unaltered. But we see that such
a situation is not realizable as a matter of principle, since we
can never exclude the possible creation of unobserved soft photons
(actually, we cannot know the soft photon content of the initial
state either). In other words, the ideal measurement that is used
to build ordinary Quantum Mechanics is excluded due to the
relativistic effects that are described by QFT. It also seems that
QFT is even less deterministic than Nonrelativistic Quantum
Mechanics, due to this underlying sort of Indetermination
Principle on the Number of Particles. In fact, the only
predictions that it allows are on probabilities and average
values. I will come back later to this important point.

In Ref.˜ \cite{QEDEPR}, I argued that this greater indetermination
allows the EPR paradox to be removed. The argument went as
follows. First, I noticed that there are two sources of
indetermination on the number of {\it real} particles in an EPR
experiment: at the production process, or at the measuring
apparatus. In Ref.˜ \cite{QEDEPR}, I have explicitly drawn some
Feynman diagrams that predict these effects. Here, just knowing
that they exist is sufficient. In our example, after the spin
$S_z(A)$ is measured to be, say, $+\hbar/2$ on particle A, there
is now way of knowing how many soft photons there are around. At
most, we could say that the rest of the world, including particle
B and an {\it undetermined} number of {\it unobserved} soft
photons, should get a definite value of the angular momentum,
$-\hbar/2$, to compensate the result obtained on A. This means
that {\it the measurement on A does not allow for any prediction
of the value of the spin (or any other conserved quantity) on B
with certainty.} One could even look for single events showing an
apparent symmetry violation (which could be important in the case
of angular momentum), due to the fact that we observe just A and
B, i.e. a part of the particles involved in the process
\cite{QEDEPR}. In any case, no mysterious action at a distance can
be {\it observed.}

Here, one could raise a subtle question: is any non locality
implied by the fact that the rest of the world (including the soft
photons) apparently ``collapses" into a $-\hbar/2$ eigenstate of
the angular momentum, after the measurement is made on particle A
\footnote{I thank Esther P\'erez for asking me this question.}?
Even though such a non locality would not be observable, the
question is relevant as a matter of principle. The complete answer
requires a deep understanding of the process of measurement in
QFT, {\it to be interpreted as the result of a succession of
elementary scattering processes}. This will allow me to correct
the wrong assumption (ii) (see several paragraphs above), that is
usually made in the treatment of the EPR paradox. As we shall see,
this will also provide a more general and complete argument
against quantum nonlocality, that could even work in the absence
of the soft photons (but, of course, they exist!).

First of all, let me recall that in QFT the scattering matrix
respects causality, and that all the conservation laws hold
locally \cite{WeinbookI,WeinbookII}. (In perturbation theory, this
means that the conservation laws are ensured by each vertex of any
Feynman graph.) Therefore, when particle A interacts with the
measuring apparatus in an EPR experiment, the angular momentum is
conserved locally: {\it only the particles of the apparatus that
come in interaction with A} (including the ones that might be
created) {\it can change their angular momenta when the spin of A
is measured!} This is a {\it prediction} of QFT.

In other words, we find that assumption (ii) was wrong, and {\it
the measurement on A is not compensated by a collapse on B, but
merely by a change of the angular momentum state of the measuring
apparatus!} After the measurement, the state of the composite
system, A + measuring apparatus (including the possible soft
photons that might appear there), has the same angular momentum
properties than before the measurement. No instantaneous effect is
implied on B or on any distant particle. Special Relativity is
saved!

Therefore, {\it when the process of measurement is interpreted
correctly in QFT, no quantum nonlocality exists!} Incidentally, by
considering the measurement as made out of scattering processes,
we are also avoiding giving the observer the magic role that
he/she had in the old formulation of Quantum Mechanics.

This also implies that we will have to renounce to the direct
applications of the EPR paradox, such as entanglement and
teleportation. However, I hope that the views that I am presenting
here might stimulate further research on the applications of QFT,
and could provide a new base for the construction of Quantum
Information Theory and Quantum Computers, that will eventually
continue to be a theoretical and practical necessity.

As a result of the previous discussion, I can safely assume that
the EPR paradox is removed in QFT. But then one could wonder: what
does the known QFT of Particle Physics actually predict for the
correlations that have been measured by now in EPR experiments? Is
there any risk of the agreement of the old Quantum Mechanics with
the data being affected? In Ref.˜ \cite{QEDEPR}, I have given a
rough computation that shows that the correlations are generally
smaller than those obtained by the usual, old ``entanglement"
theory, which were ``maximal" in our simple example. The
difference was roughly proportional to the total probability for
the appearance of an odd number of soft photons. If in the
considered experimental settings that probability is small enough,
the correlations predicted by QFT are expected to be close to the
ones that were obtained by the old, soft photons-less,
entanglement theory, and that showed an agreement with the actual
data. In other words, not only are QFTs locally realistic (as we
have seen); they can still predict correlations that can violate
Bell inequalities. Incidentally, this shows that the Bell theorem
cannot be applied to QFT.

It is worth noting that such correlations are just a reminder of
the common origin in the past of the particles, and are not due to
any mysterious action at a distance. This is a general result in
QFT, where the correlation are proven to be causal, i.e.˜ they
respect Special Relativity \cite{WeinbookI,WeinbookII}. This is
considered to be necessary for the consistency of the theory
itself, which turns out to be more intuitive (or at least less
absurd) than it was thought.

At last, Einstein, Podolsky and Rosen were right at least on one
basic principle: nonlocality is impossible. However, the solution
to their paradox does not reside in Hidden Variable Theories and
determinism, but in the Quantum Field Theories that have been
introduced to describe Particle Physics
\cite{WeinbookI,WeinbookII}, and that represent one of the most
beautiful and successful achievements in the History of Physics.
What Einstein did not expect was the fact that it was not
necessary to renounce to the indetermination to solve the paradox.
In fact, QFT seems to be even less deterministic than the old
Quantum Mechanics!

I am grateful to Humberto Michinel for stimulating discussions,
and to him and Fernando Tommasini for encouragement and for
critically reading the preliminary versions of this work. I also
thank Ruth Garc\'\i a Fern\'andez and Rebecca Ramanathan for help.

\bibliography{EPRsolve}

\end{document}